\documentclass[useAMS,usenatbib]{mn2e}

\usepackage{graphicx}
\usepackage{natbib}
\usepackage{amssymb,amsmath}
\newif\ifAMStwofonts
\AMStwofontstrue

\newcommand{\eg}{{e.g.}}
\newcommand{\gsim}{\,\lower2truept\hbox{${>\atop\hbox{\raise4truept\hbox{$\sim$}}}$}\,}

\def\eg{{\rm e.g.$\,$}}

\newcommand{\be}{\begin{equation}}
\newcommand{\ee}{\end{equation}}
\newcommand{\bea}{\begin{eqnarray}}
\newcommand{\eea}{\end{eqnarray}}

\renewcommand{\vec}[1]{ {\bmath #1} } 

\def\ltsima{$\; \buildrel < \over \sim \;$}
\def\simlt{\lower.5ex\hbox{\ltsima}}
\def\gtsima{$\; \buildrel > \over \sim \;$}
\def\simgt{\lower.5ex\hbox{\gtsima}}

  \newcommand{\bc}{\begin{center}}
  \newcommand{\ec}{\end{center}}
  
  \newcommand{\hMsun}{h^{-1}\>{\rm M_\odot}}

\title[Mass function in cDE models]{The halo mass function in interacting Dark Energy models}

\author[W.~Cui, M.~Baldi, S.~Borgani]{Weiguang Cui$^{1}$, Marco
  Baldi$^{2,3}$, Stefano Borgani$^{1,4,5}$ \\  
$^1$ Astronomy Unit, Department of Physics, University of Trieste, via
Tiepolo 11, I-34131 Trieste, Italy (wgcui,borgani@oats.inaf.it)\\
$^{2}$ Excellence Cluster Universe, Boltzmannstr.~2, D-85748
  Garching, Germany (marco.baldi@universe-cluster.de)\\
$^{3}$ University Observatory, Ludwig-Maximillians University
  Munich, Scheinerstr. 1, D-81679 Munich, Germany\\
$^4$ INAF, Astronomical Observatory of Trieste, via Tiepolo 11,
I-34131 Trieste, Italy\\ 
$^5$ INFN -- National Institute for Nuclear Physics, Trieste, Italy\\
}

\begin{document}


\pagerange{\pageref{firstpage}--\pageref{lastpage}} \pubyear{2011}

\maketitle

\label{firstpage}

\begin{abstract}	
\ \\
We present a detailed investigation of the effects that a direct 
interaction between Dark Energy and Cold Dark Matter particles imprints 
on the Halo Mass Function of groups and clusters of galaxies. Making use 
of the public halo catalogs of the {\small CoDECS} simulations, we 
derive the Halo Mass Function for several different types of coupled 
Dark Energy scenarios both based on the Friends-of-Friends algorithm 
and on the Spherical Overdensity halo identification for different 
values of the overdensity threshold $\Delta _{c}$. We compare the 
computed Halo Mass Functions for coupled Dark Energy cosmologies with 
$\Lambda $CDM as well as with the predictions of the standard analytic 
fitting functions. Our results show that the standard fitting functions 
still reproduce reasonably well both the Friends-of-Friends and the 
Spherical Overdensity Halo Mass Functions of interacting Dark Energy 
cosmologies at intermediate masses and at low redshifts, once rescaled 
to the characteristic amplitude of linear density perturbations of 
each specific model as given by $\sigma _{8}$. However, we also find that 
such apparent degeneracy with $\sigma _{8}$ is broken both by the 
high-mass tail and by the redshift evolution of our Halo Mass Functions, 
with deviations beyond $\sim 10\%$ for most of the models under 
investigation. Furthermore, the discrepancy with respect to the 
predictions of standard fitting functions rescaled with the 
characteristic value of $\sigma _{8}$ shows -- for some models -- a 
strong dependence on the spherical overdensity threshold $\Delta _{c}$ 
used for the halo identification. We find that such effect is due to a 
significant increase of halo concentration at low redshifts in these 
models, that is however absent in the majority of the cosmological 
scenarios considered in this work. We can therefore conclude that 
the universality of the Halo Mass Function is violated by cosmological 
models that feature a direct interaction between Dark Energy and 
Cold Dark Matter.
\end{abstract}

\begin{keywords}
dark energy -- dark matter --  cosmology: theory -- galaxies: formation
\end{keywords}


\section{Introduction}
\label{i}

Despite its longstanding success in explaining a wide range of
astrophysical and cosmological observations, the standard cosmological
model based on a cosmological constant $\Lambda $ as the driver of the
observed accelerated expansion \citep{Riess_etal_1998,
  Perlmutter_etal_1999,Schmidt_etal_1998} and on Cold Dark Matter
(CDM) particles as the source of structure formation processes, is far
from being accepted as a satisfactory description of the Universe. The
lack of a firm physical interpretation of the two main dark
constituents of the cosmic energy budget, Dark Energy (DE) and CDM,
represents in fact an open issue that is presently motivating an
enormous effort in the community both concerning the development of
alternative theoretical models and the implementation of observational
tests capable to falsify the standard scenario.

Furthermore, the severe fine-tuning problems that characterize the
cosmological constant \citep[see e.g.][]{Weinberg_1988} and the
increasing number of astrophysical observations suggesting a possible
tension with the predictions of the $\Lambda $CDM paradigm \citep[see
e.g.][]{BoylanKolchin_Bullock_Kaplinghat_2011,Lee_Komatsu_2010,
  Thomas_Abdalla_Lahav_2011,Lowell_etal11} have stimulated the
development of a wide variety of alternative scenarios. These range
from simple dynamical DE models based on a minimally coupled scalar
field as e.g. {\em Quintessence}
\citep[][]{Wetterich_1988,Ratra_Peebles_1988} or {\em k-essence}
\citep[][]{kessence} to interacting DE models
\citep[][]{Wetterich_1995,Amendola_2000}, to modifications of General
Relativity at cosmological scales \citep[see e.g.][]{Hu_Sawicki_2007},
Warm Dark Matter scenarios \citep[][]{Bode_Ostriker_Turok_2001}, or
Clustering DE models \citep[][]{Creminelli_etal_2010}.  All these
models should therefore be tested by a direct comparison of their
characteristic features with observations. This requires to develop
accurate predictions of the effects of all these different scenarios
on various observable quantities, from the background evolution down
to the highly nonlinear regime of cosmic structure formation

In the next decade, a large number of observational ventures, such as
the ESA approved EUCLID\footnote{\tt http://www.euclid-ec.org/} mission
\citep{euclidrb}, will enormously improve the constraints on the
minimal set of parameters that characterize the standard $\Lambda $CDM
cosmological scenario.
However, the quest for a physical interpretation of the dark components
of the Universe demands for an extension of these observational probes
beyond such minimal parameters set, allowing for a richer
phenomenology of the dark sector with respect to what envisaged by the
standard model.  In fact, although clearly necessary, the continuous
improvement of the observational constraints on the standard
cosmological parameters might not be sufficient to detect possible
failures of the $\Lambda $CDM scenario, thereby preventing a deeper
understanding of the physical properties of the dark universe. This is
mainly due to the fact that the observable deviations from the
expectations of the standard model are in many cases totally or
partially degenerate with one or more of the standard cosmological
parameters, which makes it particularly difficult to disentangle the
two effects.

It is therefore a crucial task for a full exploitation of present and
future observational probes to investigate to which extent the
characteristic features of a wide range of alternative cosmological
scenarios might be hidden beyond a simple shift in one or more of the
standard $\Lambda $CDM cosmological parameters. In particular, in the
present paper we will study in great detail the halo mass function
(hereafter HMF) of groups and clusters of galaxies in the context of a
significant number of interacting Dark Energy scenarios and we will
investigate their apparent degeneracy with the amplitude of linear
density perturbations encoded by the power spectrum normalization
$\sigma _{8}$. Indeed, a precise calibration of the HMF represents the
basic ingredient to exploit the potential of galaxy clusters as
tracers of cosmic growth \citep[see e.g.,][ for a recent
review]{Allen_etal11}.

Within the standard $\Lambda$CDM paradigm, a number of HMF studies
have been carried out through the years, based on N--body simulations
covering progressively larger dynamic ranges, with the purpose of
calibrating fitting functions for a universal HMF
\citep[e.g.][]{Sheth99,Jenkins01,Reed03,Warren06,Lukic07},
or to characterize subtle deviations from this universality
\citep[e.g.][]{Reed07,Tinker08,crocce_etal10}. Extensions of these
studies to include the effect of baryons on the HMF have been made
possible only recently by the advent of cosmological hydrodynamical
simulations covering an adequate range of halo masses
\citep{Rudd08,Stanek09,Cui_etal11}. It is however clear that the quest
of exploring cosmological models beyond the standard one, requires
enlarging the range of models for which the HMF calibration needs to
be carried out, and whose universality with respect to variations of
the parameters relevant for each specific class of models needs to be
assessed. For instance, \cite{bhattacharya_etal11} analyzed the HMF
for an extended suite of simulations of quintessence models, and
confirmed violation of universality at the $\sim 10$ per cent level
for the range of masses and redshift covered by their simulations.

A semi--analytic study of the HMF for coupled DE models has been
recently carried out by \cite{Tarrant_etal11}. Using the expressions
for a universal $\Lambda$CDM HMF, they computed the variations in the
cluster counts induced by the modification in the spherical collapse
model induced by the DE coupling to CDM. 

In this paper, we make use of the publicly available catalogs of the
{\small CoDECS}\footnote{\tt
  http://www.marcobaldi.it/web/CoDECS.html}
N-body simulations \citep[][]{CoDECS}, to analyse in detail the effect
that different coupled DE models have on the evolution of the HMF. We
will directly compare the number counts of massive halos identified
both through a Friend-of-Friend (FoF) algorithm and through a
Spherical Overdensity (SO) algorithm at different overdensities with
the standard fitting functions calculated assuming different values of
$\sigma _{8}$. Former analysis of the HMF for N--body simulations of
Coupled DE models were presented in \cite{Baldi_etal_2010} and 
\citet{Baldi_2011a}. Their
analysis, which was based on simulations covering a narrower dynamic
range and parameter space than those analysed here, and only
considered halos identified from the FoF algorithm, showed that HMF in
Coupled DE models can be described well by the $\Lambda$CDM HMF after
suitably rescaling $\sigma_8$. This implied that the signature of
coupling in the observational analysis of the cluster mass function
could be completely masked by a modified power spectrum
normalization. In our analysis we will verify whether this degeneracy
between DE coupling and power spectrum normalization can be broken at
the level of accuracy allowed by these simulations, by following the
redshift evolution of the HMF.

The paper is organized as follows. In Section 2 we describe the
Coupled DE models considered in our analysis. Section 3 is devoted to
the description of the corresponding simulations and to the halo
identification procedure. In Section 4 we will present the results of
our analysis of the HMF and their interpretation in terms of variation
of halo concentration induced by the presence of the DE coupling. We will
summarize our main results and draw our conclusions in Section 5.

\section{The Coupled Dark Energy Models}

\label{models}
\begin{table*}
\begin{center}
\begin{tabular}{llccccc}
\hline
\hline
Model & Potential & $\alpha$ & $\beta_{0}$ & $\beta_{1}$ & $w_{\phi}(z=0)$ & $\sigma _{8}(z=0)$ \\
\hline 
$\Lambda $CDM & $V(\phi ) =
A$ & -- & -- & -- & $-1.0$ & $0.809$ \\ 
EXP001 & $V(\phi ) =
Ae^{-\alpha \phi }$ & 0.08 & 0.05 & 0 & $-0.997$ & $0.825$ \\ 
EXP002 & $V(\phi ) = Ae^{-\alpha \phi }$ & 0.08 & 0.1 & 0 & $-0.995$ & $0.875$ \\ 
EXP003 & $V(\phi ) = Ae^{-\alpha \phi }$ & 0.08 & 0.15 & 0 & $-0.992$ & $0.967$ \\ 
EXP008e3 & $V(\phi ) = Ae^{-\alpha \phi }$ & 0.08 & 0.4 & 3 & $-0.982$ & $0.895$ \\ 
SUGRA003 & $V(\phi ) = A\phi^{-\alpha }e^{\phi ^{2}/2}$ & 2.15 & -0.15 & 0 & $-0.901$ & $0.806$ \\ 
\hline
\hline
\end{tabular}
\caption{The list of cosmological models considered in the {\small CoDECS}
  project and their specific parameters.}
\label{tab:models}
\end{center}
\end{table*}

Interacting DE models have been widely discussed and investigated in
the literature in the last years \citep[see
\eg][]{Amendola_2000,Amendola_2004,Farrar2007,Pettorino_Baccigalupi_2008,
  Amendola_Baldi_Wetterich_2008,Koyama_etal_2009,Honorez_etal_2010,Baldi_2011a,
  Tarrant_etal_2011,Clemson_etal_2011} and a new detailed presentation
of these cosmological scenarios would be superfluous in the present
paper. We therefore refer the interested reader to the above mentioned
literature for a thorough description of coupled DE (cDE) models and
for the derivation of their main equations. We limit our discussion
here to the definition of the notation and of the conventions adopted
for the specific cDE models considered in our analysis.

In the present paper, we will consider the set of cDE models presently
included in the {\small CoDECS} suite of N-body simulations
\citep[][]{CoDECS} -- the largest set of cosmological simulations to
date for interacting DE cosmologies -- that have been presented and
discussed in \citet{Baldi_2011c} and \citet{CoDECS}.  These are flat
cosmological models where the role of DE is played by a dynamical
scalar field $\phi $ with a self-interaction potential $V(\phi )$
exchanging energy-momentum with the CDM fluid through an interaction
term defined by the following set of equations:
\begin{eqnarray}
\label{klein_gordon}
\ddot{\phi } + 3H\dot{\phi } +\frac{dV}{d\phi } &=& \sqrt{\frac{2}{3}}\beta _{c}(\phi ) \frac{\rho _{c}}{M_{{\rm Pl}}} \,, \\
\label{continuity_cdm}
\dot{\rho }_{c} + 3H\rho _{c} &=& -\sqrt{\frac{2}{3}}\beta _{c}(\phi )\frac{\rho _{c}\dot{\phi }}{M_{{\rm Pl}}} \,, \\
\label{continuity_baryons}
\dot{\rho }_{b} + 3H\rho _{b} &=& 0 \,, \\
\label{continuity_radiation}
\dot{\rho }_{r} + 4H\rho _{r} &=& 0\,, \\
\label{friedmann}
3H^{2} &=& \frac{1}{M_{{\rm Pl}}^{2}}\left( \rho _{r} + \rho _{c} + \rho _{b} + \rho _{\phi} \right)\,,
\end{eqnarray}
where the subscripts $b$, $c$, and $r$ indicate the baryonic, CDM, and
radiation components of the universe, respectively. In
Eqs.~(\ref{klein_gordon}-\ref{friedmann}) an overdot represents a
derivative with respect to the cosmic time $t$, $H\equiv \dot{a}/a$ is
the Hubble function, and $M_{{\rm Pl}}\equiv 1/\sqrt{8\pi G}$ is the
reduced Planck mass.  The source terms at the right hand side of
Eqs.~(\ref{klein_gordon},\ref{continuity_cdm}) represent the
interaction between DE and CDM, where the dimensionless coupling
function $\beta _{c}(\phi )$ sets the coupling strength while the sign
of the quantity $\beta _{c}(\phi )\dot{\phi }$ defines the direction
of the energy-momentum flow between the two components.  The energy
exchange determines a time variation of the CDM particle mass,
according to the equation:
\begin{equation}
\label{mass}
\frac{d \ln M_{c}}{dt} = -\sqrt{\frac{2}{3}}\beta _{c}(\phi )\dot{\phi }\,,
\end{equation}
which can be derived from Eq.~(\ref{continuity_cdm}).

In the present work, we will consider two possible choices for the
coupling function $\beta _{c}(\phi )$, defined as:
\begin{equation}
\beta _{c}(\phi ) = \beta _{0}e^{\beta _{1}\phi } \,,
\end{equation}
namely a constant coupling ($\beta _{1} = 0$) and an exponentially
growing coupling ($\beta _{1} > 0$).  The latter case, first proposed
by \citet{Amendola_2004} and subsequently investigated by
\citet{Baldi_2011a}, allows for larger values of the present coupling
strength $\beta _{0}$ as compared to constant coupling models, since
the impact of the interaction on the background expansion history of
the universe and on the Cosmic Microwave Background anisotropies is
strongly suppressed by the time evolution of the scalar field $\phi $.
Furthermore, we will consider two distinct choices also for the scalar
self-interaction potential $V(\phi )$, namely an exponential potential
\citep[][]{Lucchin_Matarrese_1985,Wetterich_1988,Ferreira_Joyce_1998}:
\begin{equation}
\label{exponential}
V(\phi) = Ae^{-\alpha \phi}
\end{equation}
and a SUGRA potential \citep{Brax_Martin_1999}:
\begin{equation}
\label{SUGRA}
V(\phi) = A\phi ^{-\alpha }e^{\phi ^{2}/2} \,,
\end{equation}
where for simplicity the field $\phi $ has been expressed in units of
the reduced Planck mass in Eqs.~(\ref{exponential},\ref{SUGRA}).  The
main phenomenological difference between these two potential functions
resides in the existence of a global minimum at a finite $\phi $ value
for the SUGRA potential, while the exponential potential is
monotonically decreasing to zero for $\phi \rightarrow \infty $. The
presence of a global minimum in the SUGRA potential allows for an
inversion of the scalar field motion and for a consequent change of
sign -- in case of a constant coupling $\beta _{c}$ -- of the quantity
$\beta _{c}\dot{\phi }$, as discussed in \citet{Baldi_2011c}
\citep[see also][]{Tarrant_etal_2011}.  Due to such inversion, the DE
equation of state parameter $w_{\phi }\equiv p_{\phi }/\rho _{\phi }$
shows a ``bounce" on the cosmological constant ``barrier" $w_{\phi } =
-1$, for which this class of models has been dubbed the ``Bouncing cDE
scenario" \citep[][]{Baldi_2011c}.  For the specific model considered
in the present work, the ``bounce" happens at relatively recent
epochs, $z_{\rm inv}\approx 6.8$, and has significant consequences on
the evolution of linear and nonlinear perturbations \citep[see
again][]{Baldi_2011c}. 
The effect of the coupling on the background evolution of the universe
is to allow for a phase of Early Dark Energy which goes under the name of
$\phi $-MDE \cite[$\phi $-Matter Dominated Epoch, see][]{Amendola_2000}
or G-$\phi $-MDE \cite[Growing-$\phi $-Matter Dominated Epoch, see][]{Baldi_2011a}
for models with constant and variable couplings, respectively. Such scaling
behavior of the DE density determines a different expansion history
of cDE models with respect to a $\Lambda $CDM cosmology with the same cosmological
parameters, which represents one of the most characteristic features of cDE
scenarios and that is correctly taken into account in the numerical implementation
of the {\small CoDECS} simulations described in the next Section.
All the features and the parameters of the
different models investigated in the present work are summarized in
Table~\ref{tab:models}.
\ \\

The effect of the coupling on the growth of linear density
perturbations is described by the evolution equations for linear
fluctuations $\delta _{b,c}$ for baryons and CDM, respectively
\citep[see \eg][]{Amendola_2004,Pettorino_Baccigalupi_2008}:
\begin{eqnarray}
\label{gf_c}
\ddot{\delta }_{c} &=& -2H\left[ 1 - \beta _{c}\frac{\dot{\phi}}{H\sqrt{6}}\right] \dot{\delta }_{c} + 4\pi G \left[ \rho
  _{b}\delta _{b} + \rho _{c}\delta _{c}\Gamma _{c}\right] \,, \\
\label{gf_b}
\ddot{\delta }_{b} &=& - 2H \dot{\delta }_{b} + 4\pi G \left[ \rho _{b}\delta _{b} + \rho _{c}\delta _{c}\right]\,,
\end{eqnarray}
where for simplicity the field dependence of the coupling function
$\beta _{c}(\phi )$ has been omitted.  The factor $\Gamma _{c}$ is
defined as $\Gamma _{c}\equiv 1 + 4\beta _{c}^{2}(\phi )/3$ and
represents a fifth-force acting on CDM perturbations, while the term
$2\beta _{c}(\phi )\dot{\phi }/\sqrt{6}$ is an extra-friction arising
as a consequence of momentum conservation.  The combination of these
two effects determines the deviation of the evolution of linear
density perturbations from the standard $\Lambda $CDM case.  At the
nonlinear level, the same two extra terms appear in the acceleration
equation of coupled particles \citep[see][for a derivation of the
acceleration equation]{Baldi_etal_2010}:
\begin{equation}
\dot{\vec{v}}_{c} = \beta _{c}(\phi )\frac{\dot{\phi }}{\sqrt{6}}\vec{v}_{c} - \vec{\nabla }\left[ \sum_{c}\frac{GM_{c}(\phi )\Gamma _{c}}{r_{c}} + \sum_{b}\frac{GM_{b}}{r_{b}}\right] \,,
\end{equation}
where $r_{c,b}$ is the physical distance of the target coupled
particle from the other CDM and baryonic particles, respectively.
However, in this case the relative orientation of the particle's
velocity $\vec{v}_{c}$ and of the local gradient of the gravitational
potential $\vec{\nabla }\Phi $ introduces an additional degree of
freedom and plays an important role in the evolution of nonlinear
structures in the context of cDE scenarios \citep[see][for a detailed
discussion of nonlinear effects in cDE models]{Baldi_2011b}.

 Clearly, for any chosen coupling $\beta _{c}(\phi )$ and
  potential slope $\alpha $ in a cDE model, it is always possible to
  reconstruct an effective equation of state $w_{{\rm eff}}(z)$ for a
  minimally coupled scalar field cosmology that provides the same
  expansion history.  In this respect, the specific background
  evolution of any cDE model does not represent a ``smoking gun"
  capable to uniquely identify the presence of an interaction in the
  dark sector, as the same expansion history might be provided by a
  suitably tuned standard {\em Quintessence} or {\em phantom} DE
  fluid. On the other hand, any pair of cDE models with different
  coupling functions but with identical cosmological parameters at
  $z=0$ (except for the value of the equation of state) will
  necessarily have different background expansion histories due to
  their different $\phi$-MDE scaling solutions in matter domination
  \citep[since the energy fraction of the Early Dark Energy component
  during $\phi $-MDE is proportional to $\beta ^{2}(\phi )$, see
  e.g.][]{Amendola_2000,Baldi_2011a}.  Therefore, in order to
  investigate solely the impact of the DE-CDM coupling on the growth
  of structures one would need to compare any given cDE model with a
  specific standard {\em Quintessence} or {\em phantom} DE scenario
  having the same expansion history. However, this would necessarily
  imply to choose a different reference model for every different
  coupling function $\beta (\phi )$. In our study, instead, we are
  interested in comparing a range of different cDE models with one
  single minimally coupled reference scenario, given by the
  concordance $\Lambda $CDM cosmology, all having the same WMAP7
  parameters at $z=0$, by including all the effects (i.e. modified
  background expansion history, enhanced growth of linear density
  perturbations, particle mass variation, and scalar fifth-force).

\section{The Simulations and Halo Identification}
\label{simulations}

\subsection{The N-body simulations}
\begin{table}
\begin{center}
\begin{tabular}{cc}
\hline
Parameter & Value\\
\hline
$H_{0}$ & 70.3 km s$^{-1}$ Mpc$^{-1}$\\
$\Omega _{\rm CDM} $ & 0.226 \\
$\Omega _{\rm DE} $ & 0.729 \\
$\sigma_{8}$ & 0.809\\
$ \Omega _{b} $ & 0.0451 \\
$n_{s}$ & 0.966\\
\hline
\end{tabular}
\end{center}
\caption{Parameters of the reference $\Lambda$CDM model which ia used
  to generate initial conditions for all the cDE simulations (see text).}
\label{tab:parameters}
\end{table}

For our analysis we will make use of the public data of the {\small
  CoDECS} simulations \citep[][]{CoDECS} -- the largest suite of
cosmological simulations to date for cDE cosmologies -- which include
all the specific models listed in Table~\ref{tab:models}. 
The public data of the {\small CoDECS} project have already been
used to investigate the characteristic features of cDE scenarios in terms 
of $z$-space distortions \citep[][]{Marulli_Baldi_Moscardini_2011}, of the
expected Weak Lensing signatures \citep[][]{Beynon_etal_2011}
and of the abundance of ``Bullet-like" systems \citep[][]{Lee_Baldi_2011}.

In the present work, we will consider in particular
the {\small L-CoDECS} runs, that follow
the evolution of $1024^{3}$ CDM and $1024^{3}$ baryon particles in a
periodic cosmological box of $1$ comoving $h^{-1}$Gpc aside. The
simulations have been carried out with the modified version by
\citet{Baldi_etal_2010} of the widely used parallel TreePM N-body code
{\small GADGET} \citep[][]{gadget-2}, which implements all the
specific features of cDE models described in Sec.~\ref{models}.  The
{\small L-CoDECS} simulations have a mass resolution at $z=0$ of
$m_{c}=5.84\times 10^{10}h^{-1}$ M$_{\odot }$ for CDM and
$m_{b}=1.17\times 10^{10}h^{-1}$ M$_{\odot }$ for baryons. 

In order to cover a large dynamic range over which to study the
  effect ou coupled DE with an affordable computational cost, our
  simulations account for the different forces acting on the uncoupled
  DM and coupled baryonic matter components, without including the
  hydrodynamic description of baryons. As such, our simulations are
  designed to follow the evolution of two populations of collisonless
  particles, which feel different gravitational forces. We refer to
  the analyses by \cite{Rudd08,Stanek09,Cui_etal11} for discussions on
  the effect of hydrodynamics on the HMF.

Initial conditions for all the simulations have been generated by
perturbing a homogeneous ``{\em glass}" particle distribution
\citep[][]{White_1994,Baugh_etal_1995} according to Zel'dovich
approximation \citep[][]{Zeldovich_1970} in order to produce a
random-phase realization of the same initial matter power spectrum,
corresponding to the one computed by the public Boltzmann code {\small
  CAMB} \citep[][]{camb} for a $\Lambda $CDM cosmology with parameters
consistent with the ``WMAP7 only Maximum Likelihood" results of
\citet{wmap7}, which are listed in Table~\ref{tab:parameters}.

Since the same random seed has been adopted for the initial conditions
generation of all the {\small L-CoDECS} runs, all the different
cosmological models share exactly the same particle displacements at
$z_{\rm CMB}\approx 1100$. The initial conditions for each specific
simulation are realized by scaling such displacements to the starting
redshift of the numerical integration $z_{\rm i}=99$ with the specific
growth function obtained by numerically solving
Eqs.~(\ref{gf_c},\ref{gf_b}) for each individual cosmology.

\subsection{The Halo Identification}

The two most common methods for halo identification in N-body
simulations are based on the Friend-of-Friend (FoF) algorithm
\citep[e.g.][]{Davis85} and on the spherical overdensity (SO)
algorithm \citep[e.g.][]{Lacey94}. The FoF halo finder has only one
parameter, $b$, which defines the linking length $\lambda $ as
$\lambda \equiv b \cdot l$, where $l=n^{-1/3}$ is the mean
inter-particle separation, with $n$ the mean particle number
density. In the SO algorithm there is also only one free parameter,
namely the mean density $\Delta_c~\rho_{crit}(z)$ contained within the
sphere over which the halo mass is computed, with $\rho_{crit}(z)$
being the critical cosmic density at redshift $z$. Note that our
definition of the spherical overdensity $\Delta_c$ is given with
respect to the critical density $\rho _{crit}$ instead of the
background matter density $\rho _{mean}$, which means that our
$\Delta_c$ is $\Omega_m$ times the commonly used overdensity parameter
$\Delta$. Each of the two halo finders has its own advantages and
shortcomings \citep[see more details in][and references
therein]{Jenkins01,White01,Tinker08}. The difference of halo mass and
HMF resulting from the two methods have been discussed in several
analysis \citep[see e.g.][]{White02,Reed03,Reed07,Cohn08,More11}.

In this paper, we applied both methods to identify halos and to
compare the respective HMF for the CoDECS set of N-body simulations.
First, the standard FoF algorithm has been applied, with a linking
parameter $b=0.2$, only over the CDM particles in the simulations as
primary tracers of the matter distribution. Subsequently, each
baryonic particle has been attached to the FoF group of its closest
CDM neighbor.  The gravitationally bound substructures of each FoF
halo have also been identified by means of the {\small SUBFIND}
algorithm \citep[see more details in][]{Springel01, Dolag09}. The
position of the most bound particle of each substructure defines then
the center of the subhalo.

As a second step, the SO halo catalogues are built using a fast SO
algorithm \citep[see also][]{Cui_etal11}.  In this SO algorithm, the
centers of spherical regions are identified with the positions of the
most bound particles within the main subhalo of each FoF group, as
identified by {\small SUBFIND}.  Meanwhile, the region encompassed by
$R_{200}$ (i.e. the spherical region enclosing a mean overdensity 200
times larger than the critical density) around each FoF halo is also
checked, and all the substructures identified by {\small SUBFIND},
that lie outside this region, are considered as new independent FoF
groups. The most bound particle of each such new independent halos is
then used again as the center of a new SO region.  For each
independent halo identified with this method we progressively increase
the radius of a sphere centered on the most bound particle until a
specified mean internal density contrast $\Delta_c~{\rho}_{crit}(z)$
is reached.  The mass $M_{\Delta_c}$ within this spherical region of
radius $R_{\Delta_c}$ is then \be
M_{\Delta_c}=\frac{4}{3}{\pi}R^3_{\Delta_c} \Delta_c
{\rho}_{crit}(z)\,.  \ee

Since each halo is firstly identified starting from a FoF algorithm,
it inherits some FoF disadvantages. A well known potential problem
with FoF is that there are situations in which two halos are connected
through a bridge of particles. Our check of all the substructures
lying outside $R_{200}$ should reduce this effect, especially for the
most massive halos.  As discussed by \citet{Reed07}, this effect
becomes more important at high redshift and for poorly resolved
low-mass halos. Since our analysis mainly concentrates on very massive
halos, which we select to be resolved by at least 200
particles, we expect the bias induced by using FoF parent groups to be
relatively small.
In order to verify this on a few test cases, we also used the {\small
  SUBFIND} results to check the mass ratio of all the subhalos within
each FoF group, with respect to the main halo. Whenever a prominent
subhalo is found whose mass is at least 0.7 times the mass of the main
halo, we treat it as a distinct halo.  We verified that 
such halos are extremely rare after we remove subhalos lying outside
$R_{200}$, at least over the mass range $M \gtrsim 10^{13} \hMsun$
considered in our analysis, so our results are not influenced by this
effect.
Finally, since the groups identified by the FoF algorithm have by
definition no overlapping, we do not include in our identification of
SO halos any restriction to prevent such overlapping (see
\citealt{Tinker08} for a discussion on halo overlapping).

We do not apply the commonly used FoF mass correction of
\cite{Warren06} for our FoF halos in this paper. This correction is
meant to regulate the halo mass, which is overestimated by the FoF
algorithm when halos are sampled with too few particles. This
correction is expected to be small for our selected halo sample,
due to the minimum threshold of 200
particles per halo that we adopted.

\section{Results}

\begin{figure*}
\includegraphics[width=1.0\textwidth]{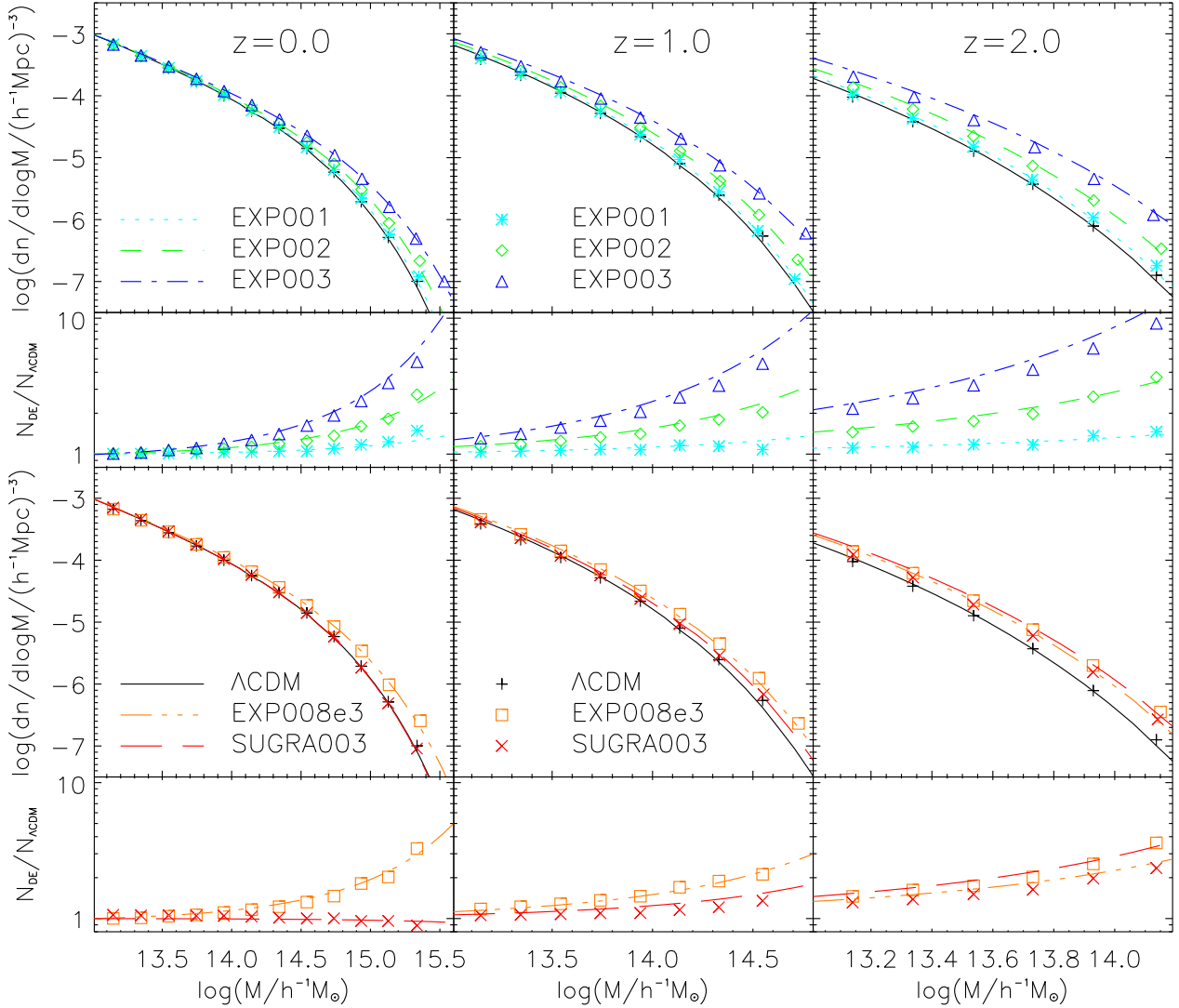}
\caption{The FoF halo mass function (HMF) for the different cDE
  simulations at $z=0$, 1 and 2 (left, central and right panels,
  respectively). In all cases $b=0.2$ is assumed for the linking
  parameter used in the FoF halo identification. In the upper part of
  each panel the symbols show the results from our simulations, while
  curves are the corresponding predictions from the FoF HMF of
  Eq. \protect\ref{eq:fitJ} calibrated by \protect\cite{Jenkins01} for
  a $\Lambda$CDM cosmology and rescaled to the value of $\sigma_8$
  appropriate for each cDE model. In all panels, black crosses and
  black curves are for the results of our $\Lambda$CDM reference
  simulation. The lower part of each panel shows the difference with
  respect to $\Lambda$CDM results. Here symbols show the difference
  between cDE and $\Lambda$CDM simulations, while curves with
  different line-styles show the difference between the corresponding
  fitting functions.}
\label{fig:HMF_FOF}
\end{figure*}

\begin{figure*}
\includegraphics[width=1.0\textwidth]{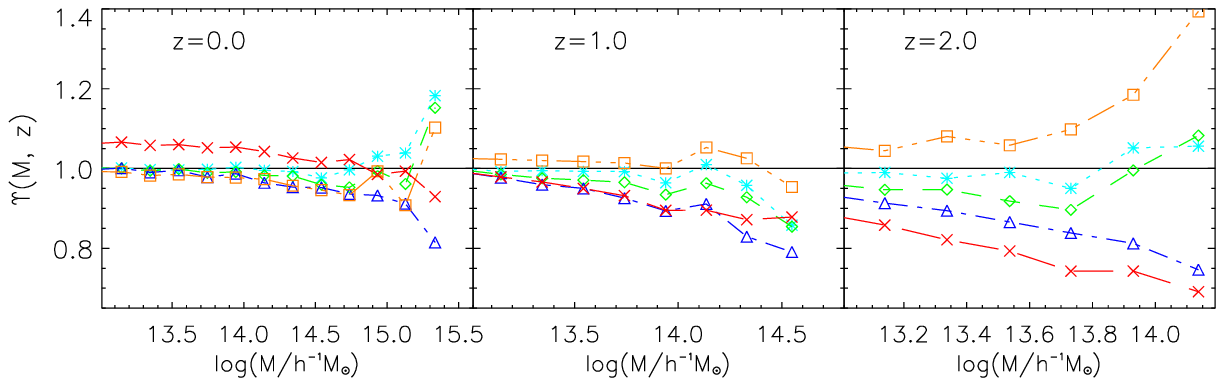}
\caption{The ratio between the FoF halo counts in different mass
  intervals from the cDE models and from the reference $\Lambda$CDM
  model from simulations, compared to the same quantity computed from
  the corresponding fitting functions from Eq. \protect\ref{eq:comp}
  (see also text). Deviations of this quantity from unity indicates
  that simply rescaling $\sigma_8$ in the fitting function of
  Eq.\protect\ref{eq:fitJ} does not correctly account for the
  difference between cDE and $\Lambda$CDM simulation results. The
  correspondence of simulated models with symbols and line-types is
  the same as in Fig. \protect\ref{fig:HMF_FOF}.}
\label{fig:HMF_D_FOF}
\end{figure*}

Under simplifying assumptions, \cite{Press74} were the first to
provide a theoretical description for the abundance of collapsed halos
as a function of their mass.  Thanks to the calibration on N-body
simulations covering wide dynamic ranges, the HMF for the standard
$\Lambda$CDM cosmology can now be accurately described by analytical
models and fitting functions, as provided by
e.g. \citet{Sheth99,Jenkins01,Reed03,Warren06,Tinker08}.  In most
applications, the HMF is expressed in terms of the so--called
multiplicity function
by: 
\be f(\sigma,z) = \frac{M}{\rho_0(z)} \frac{dn(M,z)}{d \ln
  \sigma^{-1}(M,z)},
\label{eq:TMF}
\ee
where $n(M,z)$ is the number density of halos with mass $M$ at
redshift $z$ and $\rho_0(z)$ is the background density at redshift
$z$.  In Eq.~(\ref{eq:TMF}), $\sigma^2(M,z)$ is the variance of the
linear density field, extrapolated to the redshift $z$ at which halos
are identified, which is given by
\be
\sigma^2(M,z) = \frac{D^2(z)}{2\pi^2}\int\limits_0^{\infty}k^2P(k)W^2(k,M)dk\,.
\label{eq:sigma2}
\ee
In Eq.~(\ref{eq:sigma2}), $D(z)$ is the growth factor of linear
density perturbations, $P(k)$ is the linear power spectrum of density
perturbations and $W(k,M)$ is the Fourier-space representation of a
real-space top-hat filter of radius $R$, which on average encloses a
mass $M = 4 \pi R^3 \rho_0(z) /3$, given by:
\be
W(k,R) = \frac{3}{(kR)^3} [ \sin(kR) - kR\cos(kR) ]\,.  
\ee
Using a series of N-body simulations under the standard $\Lambda$CDM
cosmology and applying the FoF algorithm with linking parameter $b =
0.2$, \citet{Jenkins01} (J01 hereafter) provided a simple fitting formula as a
function of the variance $\sigma$, which is independent of redshift:
\be 
f(\sigma) = 0.315\,\exp(-|\ln\sigma^{-1} + 0.61|^{3.8}).
\label{eq:fitJ}
\ee 
This fitting function is valid over the range $-1.2 \leq \ln
\sigma^{-1} \leq 1.05$, which corresponds to halo masses $10.3 \leq
\log (M/\hMsun) \leq 15.6$ in our $\Lambda$CDM simulation.  We use
this formula as our benchmark reference fit for the FoF HMF.  We adopt
instead the fitting function of \cite{Tinker08} (T08 hereafter) for
the SO HMF. This fitting function $f(\sigma)$ is tuned by the four
parameters $A, a, b, c$ according to:
\be f(\sigma) = A \left[\left(\frac{\sigma}{b}\right)^{-a} + 1\right]
e^{-c/\sigma^2}.
\label{eq:fitT}
\ee
These four parameters are functions of two independent variables,
namely the redshift $z$ and the overdensity $\Delta$ (for more
details, see Eqs. 5--7 and B1--B4 in T08).  As already
mentioned, T08 adopted an overdensity $\Delta $ computed
with respect to the background matter density.  In order to use this
fitting function with our definition of the overdensity $\Delta _{c}$
(i.e. with respect to the critical density of the Universe), we simply
change $\Delta $ to $\Delta_c$ in the functional dependence of the
four parameters $(A,a,b,c)$ just multiplying it by $\Omega _{m}(z)$.

Throughout this paper, we adopt the differential HMF $dn/d\log M$ to
make comparisons between theoretical predictions and simulation
results. With a simple transformation of Eq.~(\ref{eq:TMF}), the
theoretical expression for $dn/d\log M$ is given by:
\be
  \frac{dn}{d\log M} = \frac{f(\sigma)\rho_0}{M} \frac{d\ln \sigma^{-1}}{d \log M}\,,
\label{eq:TDMF}
\ee
where $f(\sigma)$ is given by Eqs.~(\ref{eq:fitJ}) and
(\ref{eq:fitT}).
 
In order to compute the differential HMF $dn/d\log M$ in a simulation
of volume $L^3$, one needs to measure the number of halos $\Delta N$
in a given logarithm mass bin $\Delta \log M$ :
\be
\frac{dn}{d\log M} = \frac{M}{L^3} \frac{\Delta N}{\Delta \log M},
\ee
where we assigned the characteristic mass $M$ for a given mass bin as
the mean mass computed over all halos belonging to that bin. For our
halo sample, we used narrow mass bins with $\Delta \log M = 0.2$.
\citet{Lukic07} had studied the error introduced by binning data, and
showed that this error is negligible with such a narrow bin.  However,
the limited statistics of high--mass halos within the simulation box
makes the determination of the HMF quite noisy in the high mass end,
especially when marrow mass bins are used.
To reduce this effect, we merge mass bins containing less than 20
objects into the adjacent lower mass bin. Each mass bin is then
weighted proportionally to the number of clusters it contains. Due to
this specific treatment of the mass bins at large masses, the bin
width can be different for different simulations. Therefore, when we
compare the number of objects in such last bins, we rescale the
cluster counts within each of them to the bin width in units of
$\Delta \log M = 0.2$.

\subsection{The FoF HMF in coupled dark energy models}

We show the FoF HMF of the different cDE models included in the
{\small CoDECS project} derived from the {\small L-CoDECS} simulations
in the upper panels of Fig.~\ref{fig:HMF_FOF}. Here, the symbols show
the results from simulations, while the different line--styles
represent the HMF predicted by the J01 fitting function
given by Eq.~(\ref{eq:fitJ}) for $\Lambda $CDM
cosmologies with the same $\sigma _{8}$ value as the different cDE
realizations (see Table~\ref{tab:models}).  In the lower panels of each plot
of Fig.~\ref{fig:HMF_FOF} we show with symbols the ratio of the halo
number density of the various cDE simulations over the corresponding
results for the standard $\Lambda$CDM simulation. In the same panels,
the curves show the same ratio, but obtained using the model
predictions from the HMF fitting functions, each computed for the
appropriate value of $\sigma_{8}$. In fact, since the HMF is
described only by the variance of the linear density field
$\sigma^2(M,z)$, the HMF fitting functions of the $\Lambda$CDM
cosmology are often assumed to predict also the HMF of other
non-standard models by a simple renormalization of the linear
perturbations amplitude $\sigma_8$.

At a first sight, by looking at Fig.~\ref{fig:HMF_FOF} such assumption
holds also for cDE models considered in the present paper, with the
predicted HMFs for the $\sigma _{8}$ values computed for each model
through linear perturbations theory fitting the cDE simulations
results reasonably well, at least at $z=0$ (left panels).  This is
true for both the standard cDE models with constant coupling, EXP001-3
(upper plots), and for the other two cDE models with different
potentials or coupling functions, EXP008e3 and SUGRA003 (lower
plots). Quite interestingly, the accuracy of such rescaling degrades
at increasing redshift, with deviations appearing at $z=1$ (central
panels), which becomes more apparent at $z=2$ (right panels).

In order to better appreciate the difference between results from cDE
simulations and fitting functions, we introduce a parameter $\Upsilon
(M, z)$, which is expressed as
\be \Upsilon (M, z) =
\frac{N_{DE}^{sim}(M, z) / N_{\Lambda CDM}^{sim}(M,
  z)}{N_{DE}^{fit}(M, z) / N_{\Lambda CDM}^{fit}(M, z)}\,,
\label{eq:comp}
\ee
where $N(M,z)$ is the number density of halos within the mass bin $M$
at redshift $z$, and the lower index of $N$ indicates the model, while
the upper index indicates either the theoretical ($fit$) or the
simulation ($sim$) results. With this definition, the quantity
$\Upsilon (M, z)$ describes the difference between simulations and
fitting functions, after rescaling each to the corresponding
$\Lambda$CDM prediction. In this way, we account for sampling effects,
due to the limited halo statistics especially in the high--end of the
simulated HMF, and for possible inaccuracies in the fitting
functions. Accordingly, a significant deviation from unity of the
$\Upsilon(M, z)$ parameter indicates a lack of precision in the
rescaling of the HMF fitting function for the cDE models.

In Figure \ref{fig:HMF_D_FOF}, we show the mass dependence of
$\Upsilon(M, z)$ at the same redshifts considered in
Fig. \ref{fig:HMF_FOF}, for the different cDE models.  At $z = 0$,
$\Upsilon$ is very close to unity at small halo masses, thus implying
that the J01 fitting function can be used to capture the
simulation results quite accurately. The only exception is represented
by the ``Bouncing cDE model" SUGRA003 (red line), that shows a $\sim
5$ per cent deviation at the low-mass end of our HMF, with a
decreasing trend at higher masses.  Deviations from unity of the
$\Upsilon$ parameter for this model change their sign at higher
redshift, where they become progressively larger. At $z=2$ simulation
results for the SUGRA003 model show a deviation from $\Lambda$CDM
simulation results that is $\sim 30$ per cent smaller than predicted
by the corresponding J01 fitting function, for the
highest sampled halos masses. An opposite trend is instead found for
the variable--coupling model EXP008e3, with $\Upsilon\simeq 1.4$ for
the highest masses sampled at $z=2$. As for the standard models with
constant coupling (EXP001-3), the value of $\Upsilon$ shows smaller
deviations from unity, which also in this case increases with mass and
redshift. As expected, the smallest deviations are found for the
models with the smallest value of the $\beta_0$ coupling (EXP001).

In general, Figure~\ref{fig:HMF_D_FOF} clearly shows that, even for
the FoF HMF, the apparent degeneracy between the coupling and $\sigma
_{8}$, which holds to first approximation in the low-mass end at low
redshift, is broken in the high-mass end by an amount which increases
with redshift, with the strength of the coupling, and whose sign also
depends on the shape of self--interaction potential for the dynamical
scalar field which determines the coupling.

\subsection{The SO HMF in coupled dark energy models}

The FoF algorithm does not assume any geometry for the halo, unlike
the enforced sphericity of SO. For this reason, the mass function for
halos identified with a SO algorithms can be more directly compared
with the observed mass function of galaxy clusters and groups, whose
mass is generally measured, or inferred from mass proxies, within some
radius encompassing a fixed overdensity $\Delta_c$ with respect to the
cosmic critical density.  In this section, we investigate the SO HMF
at three reference overdensities $\Delta_c = 200, 500, 1500$.

\begin{figure*}
\includegraphics[width=1.0\textwidth]{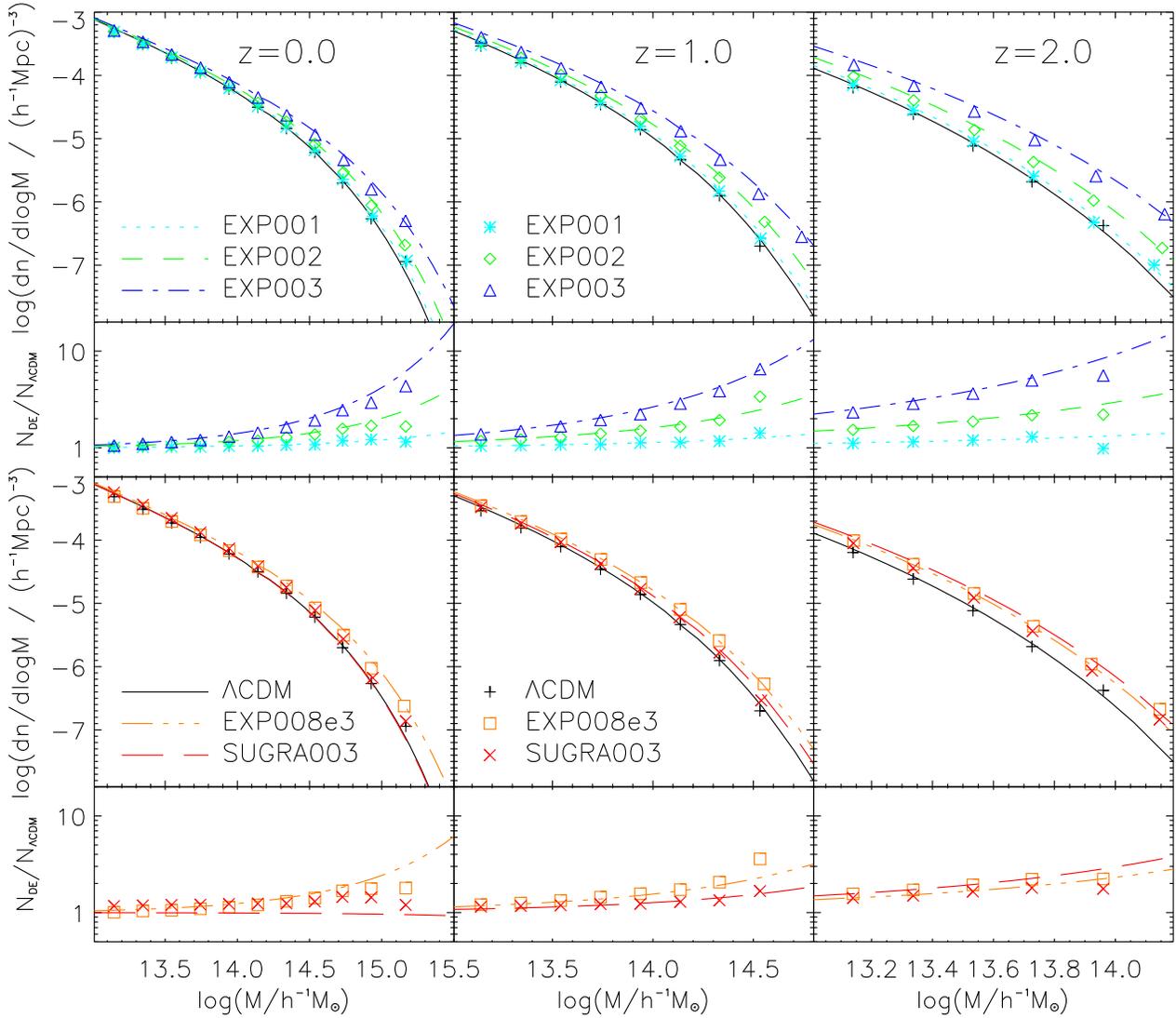}
\caption{The same as in Figure \protect\ref{fig:HMF_FOF}, but for
  the HMF computed with the spherical overdensity (SO) algorithm at
  $\Delta_c = 200$. In this case, curves are obtained from the fitting
  function of Eq. \protect\ref{eq:fitT} by \protect\cite{Tinker08},
  computed for the appropriate overdensity.}
\label{fig:HMF_SO_200}
\end{figure*}

\begin{figure*}
\includegraphics[width=1.0\textwidth]{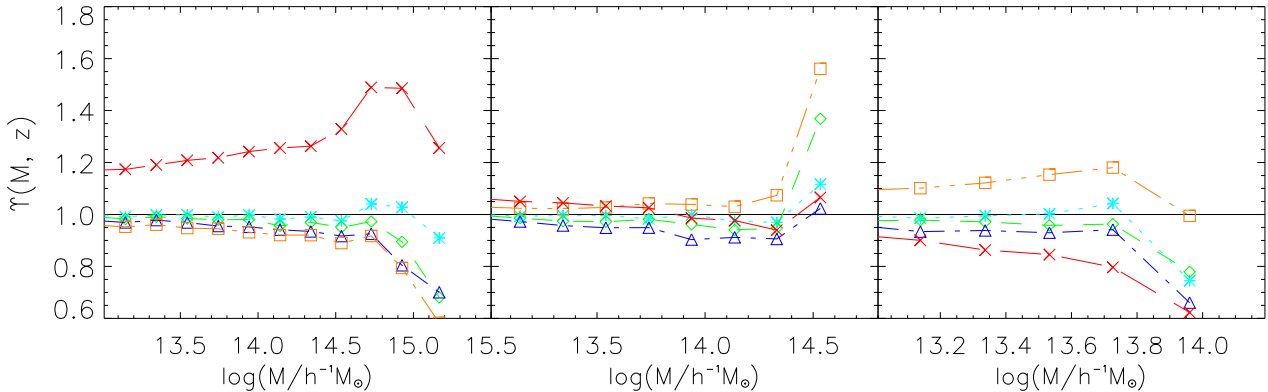}
\caption{The same as in Figure \protect\ref{fig:HMF_D_FOF}, but for
  the HMF computed with the spherical overdensity (SO) algorithm at
  $\Delta_c = 200$. In this case, curves are obtained from the fitting
  function of Eq. \protect\ref{eq:fitT} by \protect\cite{Tinker08},
  computed for the appropriate overdensity.}
\label{fig:HMF_SO_D_200}
\end{figure*}

In Figure~\ref{fig:HMF_SO_200} we show the same SO HMF properties at
$\Delta_c = 200$ as shown for the FoF HMF in Figure \ref{fig:HMF_FOF}
above.  From the upper part of the panels of Figure
\ref{fig:HMF_SO_200}, we see that at a first sight the SO HMF for
different cDE models can also be described by a simple renormalization
of $\sigma_8$ in the T08 fitting function of Eq. \ref{eq:fitT}. The
ratios to $\Lambda$CDM in the lower panels of
Figure~\ref{fig:HMF_SO_200} are also qualitatively similar to what
already shown for the FoF HMF in Figure \ref{fig:HMF_FOF}.  All the
cDE models based on an exponential potential present more massive
halos than $\Lambda$CDM at all the three redshifts considered, while
the ``Bouncing cDE model" SUGRA003 is consistent with $\Lambda$CDM at
$z=0$, but shows an excess of massive halos at higher redshifts, in
line with the results discussed by \citet{Baldi_2011c}.  Also in this
case, we show the $\Upsilon$ parameter as a function of halo mass in
Figure \ref{fig:HMF_SO_D_200}.  Differently from what shown for the
FoF halos in Figure \ref{fig:HMF_D_FOF}, the $\Upsilon$ parameter of
the three standard cDE models (EXP001--3) are quite close to unity
with a decline $< 10$ per cent at all redshifts and at all masses
(except at the very high mass end).  The $\Upsilon$ parameter for the
EXP008e3 model shows an increase with redshift, which is in any case
still within $\sim 10$ per cent.  On the other hand, the $\Upsilon$
parameter for the SUGRA003 model is above unity, by $\sim 20$--40 per
cent, at $z=0$. Its value is larger than what shown for the FoF case of
Fig. \ref{fig:HMF_D_FOF}, while dropping below unity at $z=2$ by $\sim
10$--20 per cent. In general, these results suggest that the
$\sigma_8$ rescaling applied to the T08 fitting function provides a
better fit to the SO HMF at $\Delta_c=200$ for the standard cDE models
than the same rescaling does when applied to the
J01 fitting expression for the FoF HMF, except for the bouncing cDE model SUGRA003.

\begin{figure*}
\includegraphics[width=1.0\textwidth]{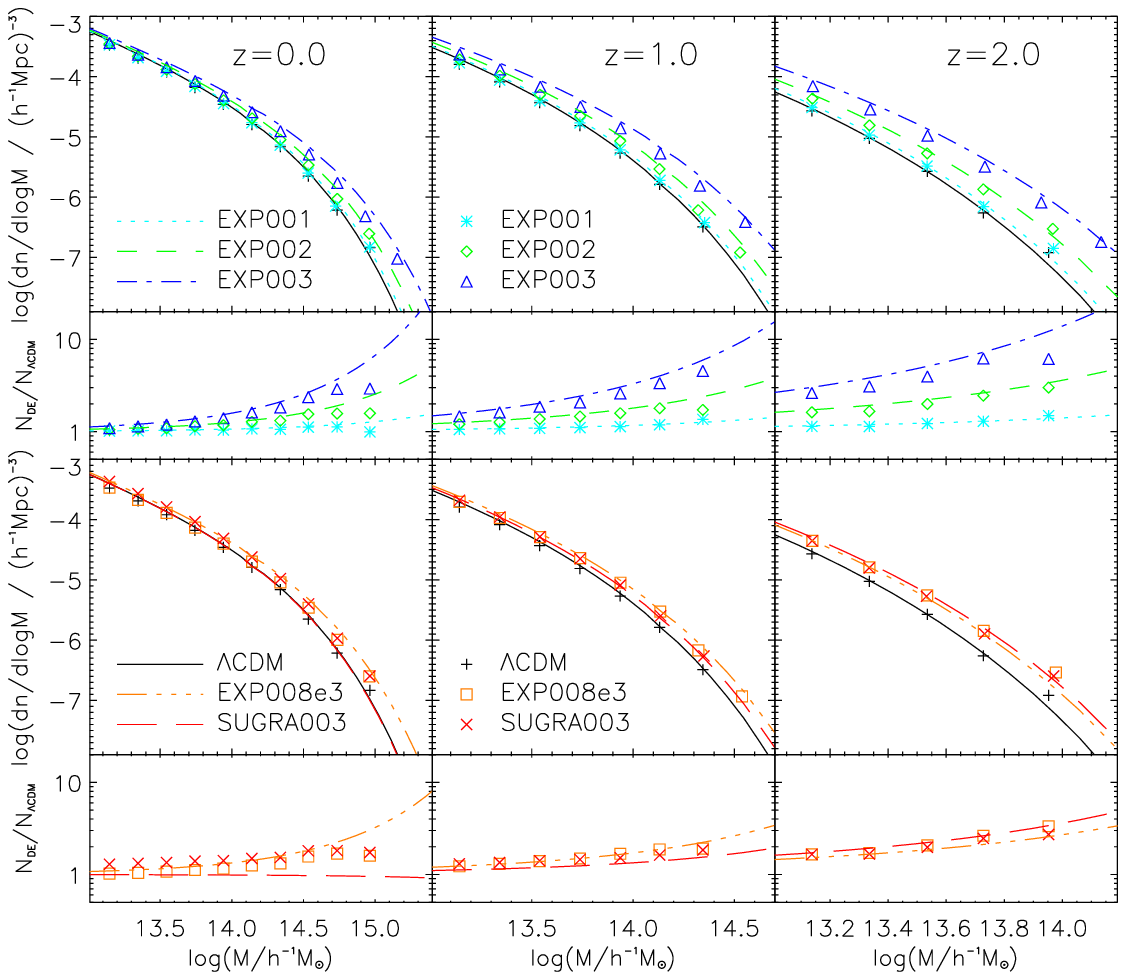}
\caption{The same as in Figure \protect\ref{fig:HMF_FOF}, but for
  the HMF computed with the spherical overdensity (SO) algorithm at
  $\Delta_c = 500$. In this case, curves are obtained from the fitting
  function of Eq. \protect\ref{eq:fitT} by \protect\cite{Tinker08},
  computed for the appropriate overdensity.}
\label{fig:HMF_SO_500}
\end{figure*}

\begin{figure*}
\includegraphics[width=1.0\textwidth]{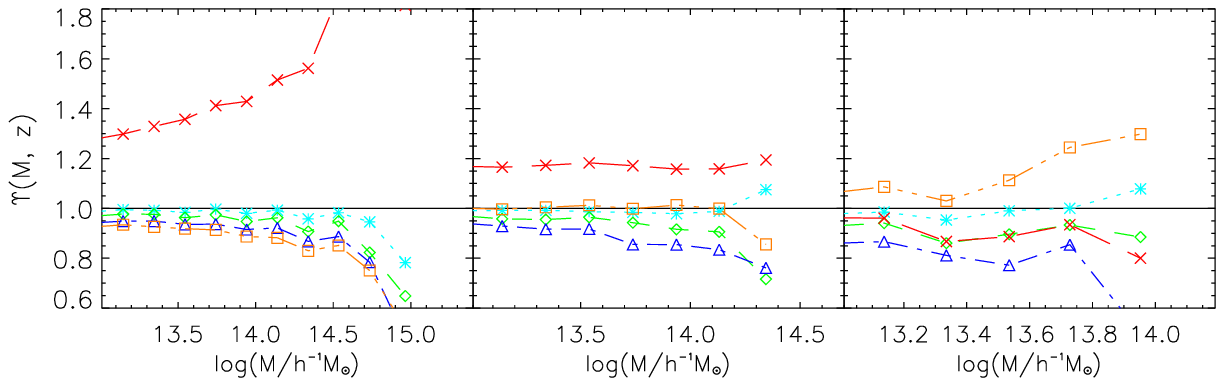}
\caption{The same as in Figure \protect\ref{fig:HMF_D_FOF}, but for
  the HMF computed with the spherical overdensity (SO) algorithm at
  $\Delta_c = 500$. In this case, curves are obtained from the fitting
  function of Eq. \protect\ref{eq:fitT} by \protect\cite{Tinker08},
  computed for the appropriate overdensity.}
\label{fig:HMF_SO_D_500}
\end{figure*}

The SO HMFs at $\Delta_c = 500$ are shown in Figure
\ref{fig:HMF_SO_500}. At this higher overdensity, the SO HMFs from
simulations are still fit by the theoretical predictions as in Figure
\ref{fig:HMF_SO_200}, with the ratio displayed in the lower panels
also showing a similar quality of the fit as for the previous
cases. From Figure \ref{fig:HMF_SO_D_500} we see that the $\Upsilon$
parameter for the standard cDE models has larger deviations from unity
than for the $\Delta_c = 200$ case, especially at higher redshift. As
for the SUGRA003 model, this parameter is confirmed to have a
decreasing trend with redshift, starting with a value at $z=0$ which
is larger than the corresponding one found for $\Delta_c = 200$.

\begin{figure*}
\includegraphics[width=1.0\textwidth]{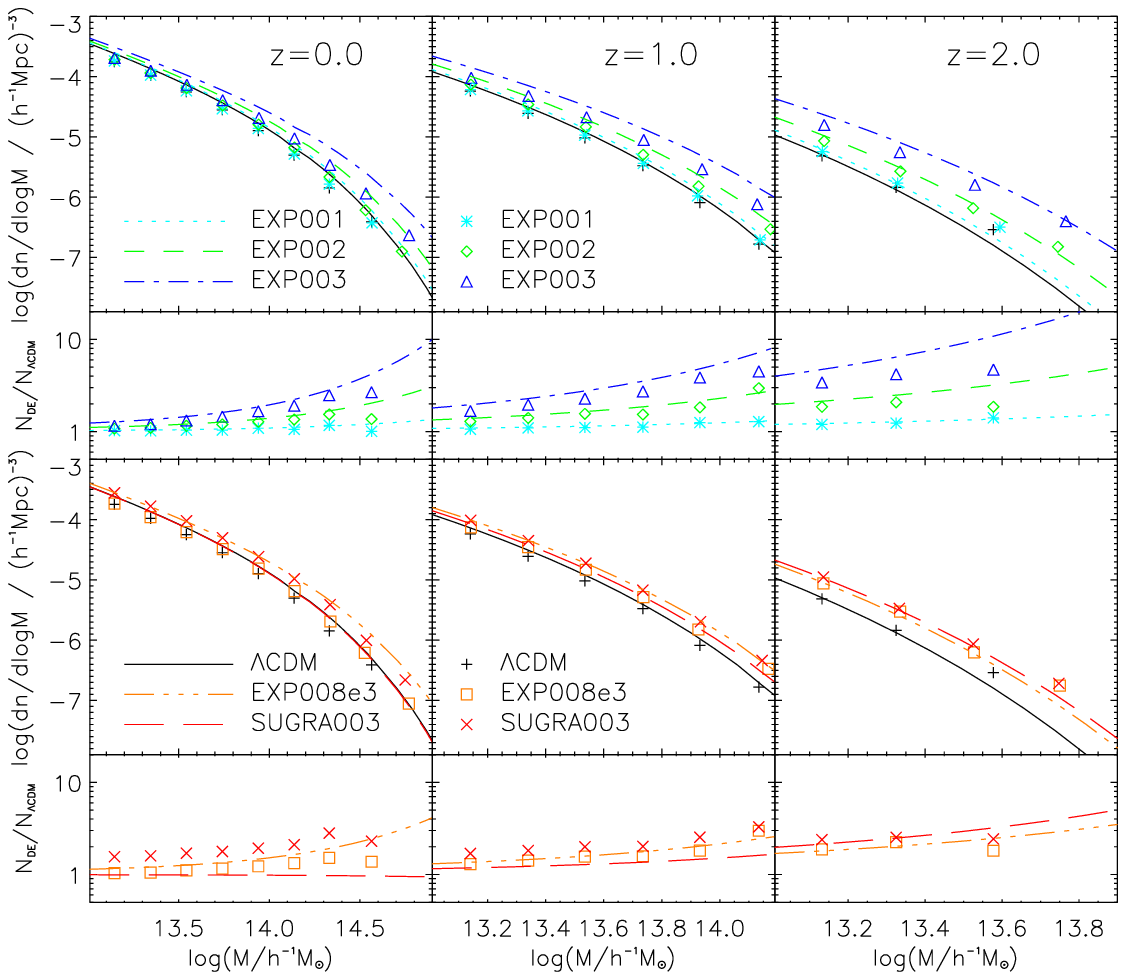}
\caption{The same as in Figure \protect\ref{fig:HMF_FOF}, but for
  the HMF computed with the spherical overdensity (SO) algorithm at
  $\Delta_c = 1500$. In this case, curves are obtained from the fitting
  function of Eq. \protect\ref{eq:fitT} by \protect\cite{Tinker08},
  computed for the appropriate overdensity.}
\label{fig:HMF_SO_1500}
\end{figure*}

\begin{figure*}
\includegraphics[width=1.0\textwidth]{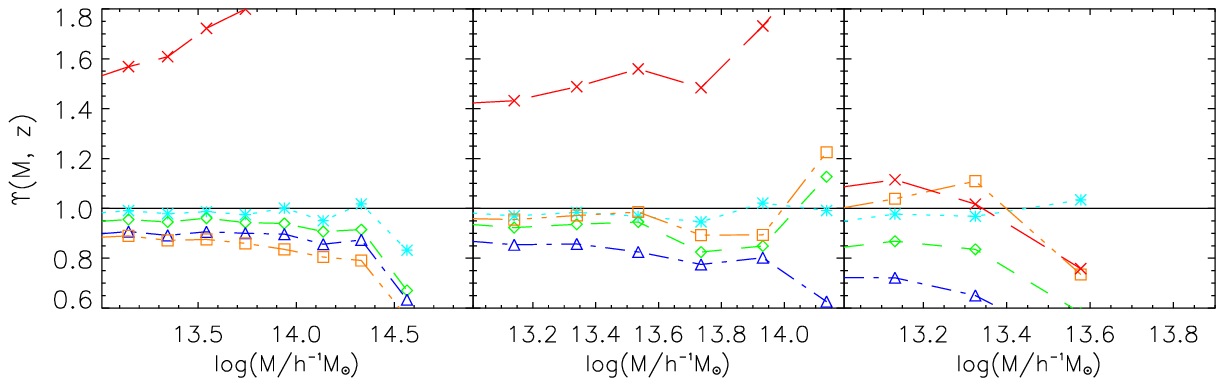}
\caption{The same as in Figure \protect\ref{fig:HMF_D_FOF}, but for
  the HMF computed with the spherical overdensity (SO) algorithm at
  $\Delta_c = 1500$. In this case, curves are obtained from the fitting
  function of Eq. \protect\ref{eq:fitT} by \protect\cite{Tinker08},
  computed for the appropriate overdensity.}
\label{fig:HMF_SO_D_1500}
\end{figure*}

In general, all these trends are amplified as we consider
progressively larger values of $\Delta_c$. In fact, at our highest
overdensity $\Delta_c = 1500$, the fitting function by T08
does not provide a good fit to the simulation results anymore, even
for the $\Lambda$CDM simulation as shown in Figure
\ref{fig:HMF_SO_1500}. Nevertheless, the ratios to corresponding
$\Lambda$CDM results have a similar variation with halo mass as in the
previous figures.  Since the $\Upsilon$ parameter is the ratio between
simulation and fitting--function results, rescaled to the
corresponding $\Lambda$CDM predictions, it should allow to remove this
offset.  Figure \ref{fig:HMF_SO_D_1500} indicates that this is indeed
the case, with the value of $\Upsilon$ still being within $\sim 20$
per cent from unity at $z=0$ for all the exponential potential models,
while at $z=2$ we find deviations from unity of the order of $\sim 30$
per cent for the most extreme EXP003 model. On the contrary, for the
SUGRA003 model $\Upsilon$ increases above unity by more than 60 per
cent at $z=0$, and drops to $\sim 1$ at $z=2$.
Some deviation from a universal shape of the HMF has been detected
also for uncoupled DE scenarios based on a classical scalar field with no
direct interactions to matter, as discussed by \eg \citet{Courtin_etal_2011}.
However, the higher amplitude of the deviation and its peculiar mass dependence
-- with the effect being progressively more pronounced for halos of higher masses --
make the imprint of cDE models on the HMF quite peculiar and potentially distinguishable from
the one of an uncoupled model.

As a general result from the analysis of the SO HMF, we point out that
rescaling the mass function in cDE models with the $\sigma_8$ value
produces results that, at least for $\Delta_c=200$, are more accurate
than for the FoF HMF. However, this rescaling does not provide
accurate predictions of the cDE HMF at higher overdensity and larger
redshift. These results highlight that degeneracy between a
non--vanishing coupling and $\sigma_8$ can in principle be broken by
observational measurements of the cluster mass function over a
sufficiently large mass and redshift range.

\subsection{Halo Concentration}
\begin{figure*}
\includegraphics[width=1.0\textwidth]{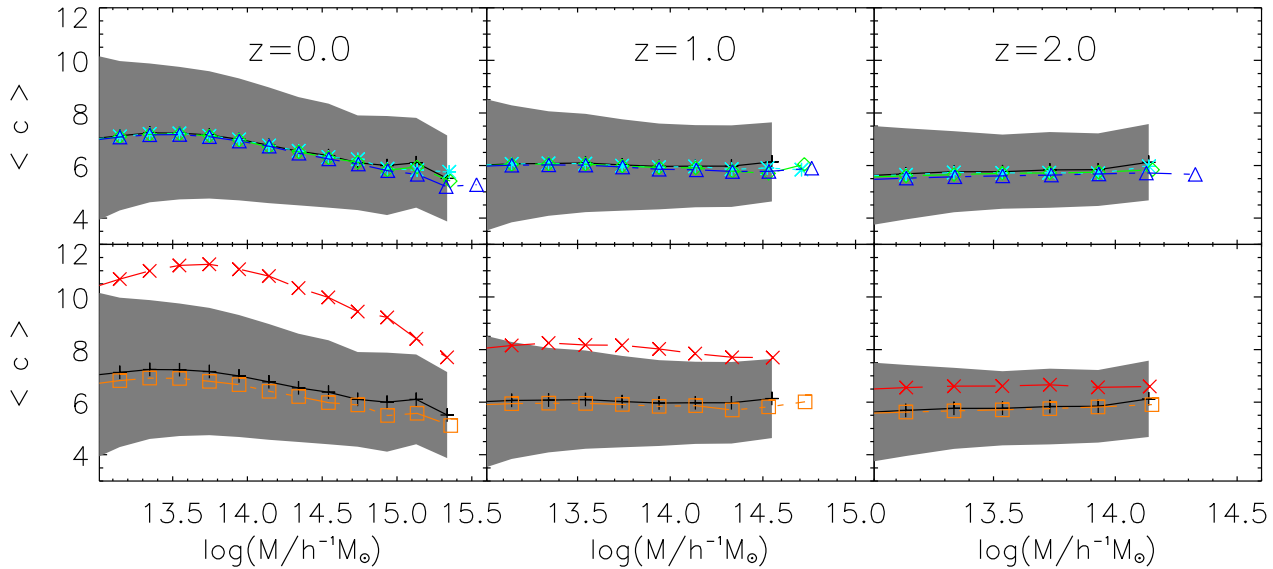}
\caption{The mean halo concentration as a function of FoF halo
  mass. The different colors, symbols and line-styles are the same as
  in the previous figures. The shaded region shows 1$\sigma$ scatter
  in the concentration computed within each mass bin. For reasons of
  clarity it is only shown for the reference $\Lambda$CDM simulation.}
\label{fig:CC}
\end{figure*}

As shown in the previous section, deviations of the SO HMF for cDE
models from the $\Lambda$CDM one increase at higher overdensity. This
result suggests that this difference in the HMF is due to a change in
the timing of halo collapse induced by the presence of coupling which,
in turn, induces a change of the halo density profiles.
 
Halos in dissipationless N-body simulations of a $\Lambda$CDM model
have spherically averaged density profiles that are well described by
the profiles of \citet{NFW97}, hereafter NFW. For an NFW halo of a
given mass, the halo density profile can be specified entirely by one
parameter, the concentration, although the relation between
concentration and mass is characterized by a significant intrinsic
scatter
\citep[e.g.][]{Maccio_etal08,Gao_etal08,Duffy_etal08,Zhao_etal09,Prada_etal11}.

In the following, we use a simple method \cite[see Eqs.~6-9
in][]{Aquarius} to compute concentrations for the halos identified in
our {\small L-CoDECS} simulations. According to this method, the
characteristic NFW overdensity $\delta_c$ is a function of halo
concentration $c$, and can be expressed in terms of the maximum
circular velocity of the halo, $V_{max}$, and the radius $r_{max}$ at
which this velocity is attained:
\be
\delta_c = \frac{200}{3} \frac{c^3}{\ln(1+c) - c/(1+c)} = 14.426 \left(\frac{V_{max}}{H_0 r_{max}}\right)^2, 
\label{eq:cc}
\ee

In Fig.~\ref{fig:CC}, we show the mean concentration $c$ calculated
from Eq.~(\ref{eq:cc}), as a function of the FoF halo mass. All the
cDE models appear to be consistent with the $\Lambda$CDM results, with
the only exception being represented by the SUGRA003 model. For this
model the average concentration is a factor of $\sim 2$ larger than in
$\Lambda$CDM at $z=0$, while gradually approaching the $\Lambda$CDM
relation at high redshift. This enhanced halo concentration in the
SUGRA003 model could provide a way to account for the apparent
over-concentration of massive clusters recently found from lensing
analyses \citep[e.g.][]{Oguri_etal09,Umetsu_etal11}. On the other
hand, if such a high concentration would hold also for typical CDM
halos of luminous spiral galaxies (a few $\times 10^{12}h^{-1}
M_{\odot}$), which are not resolved by the simulations analysed here,
this specific realization of the ``Bouncing cDE" scenario 
could run into strong tension with the observed
dynamical properties of spiral galaxies, as recently investigated by
\citet{Baldi_Salucci_2011}.  The enhancement of halo concentration at
low redshifts for the ``Bouncing cDE'' scenario also explains why the HMF
of this model shows deviations from $\Lambda $CDM only at high
overdensity thresholds, $\Delta _{c}=500\,, 1500$, while the FoF HMF
and the low overdensity SO HMF (i.e. for $\Delta _{c} = 200$) is much
closer to the expectation from the fiducial $\Lambda $CDM model.

\section{Discussion and conclusion} 
\label{sec:5}
We have investigated in large detail how the halo mass function is
affected by different possible types of interaction between dark
energy and dark matter. By means of the {\small L-CoDECS} suite of
large N-body simulations \citep{Baldi_2011c} we have computed the
abundance of halos as a function of redshift and mass in six different
cosmological models (see Table 1): the fiducial standard $\Lambda $CDM
scenario, three models of interacting DE with an exponential
self-interaction potential and a constant coupling function
(EXP001-003), one model with an exponentially growing coupling
strength (EXP008e3), and one specific realization of the Bouncing cDE
scenario based on a SUGRA self-interaction potential (SUGRA003).
  In doing so, we have assumed as a single reference scenario the
  standard concordance $\Lambda $CDM cosmology with cosmological
  parameters consistent with the latest WMAP7 results, and we have
  constructed all the other cDE models such that they have the same
  WMAP7 parameters at $z=0$. This strategy allows us to perform a
  self-consistent comparison of the footprints of cDE models with
  respect to the standard $\Lambda $CDM cosmology. On the other hand,
  in order to distinguish the peculiar signatures of a DE-CDM
  interaction on the growth of structures, from the effects of any
  other minimally coupled dynamical DE model (such as a {\em
    Quintessence} or a {\em phantom} scalar field) one would need to
  compare any given cDE scenario to a corresponding dynamical DE
  cosmology tuned to have the same expansion history.  For each of
  our scenarios, we have then derived both the
Friends-of-Friends (FoF) halo mass function and a series of
SO mass functions for three values of the overdensity
threshold $\Delta _{c}$, namely $\Delta _{c}=200\,, 500\,, 1500$, at
the present time and at $z=1$ and $2$.

Our findings show that the FoF mass function is significantly affected
by the interaction and displays large deviations from the expected
$\Lambda $CDM halo abundance at $z=0$, due to the faster growth of
density perturbations, with the only exception of the Bouncing cDE
scenario that shows very little differences from $\Lambda $CDM at the
present time. This is expected due to the peculiar dynamics of the
Bouncing cDE model that provides the same value of $\sigma _{8}$ as
$\Lambda $CDM at $z=0$. In this respect, our results fully confirm
previous findings.  On the other hand, all the remaining cDE models
are characterized by larger values of $\sigma _{8}$ as compared to
$\Lambda $CDM, and our investigation shows that the effect of the
interaction on the mass function is in general highly degenerate with
$\sigma _{8}$, such that the nonlinear halo mass function extracted
from the simulations can be quite accurately reproduced by the
standard fitting functions computed for the actual value of $\sigma
_{8}$ attained by each model. Such degeneracy is however broken by the
different redshift evolution of the halo mass function, especially at
large masses, in the various cDE models as compared to $\Lambda $CDM,
such that the same value of $\sigma _{8}$ does not accurately
reproduce the abundance of massive halos at different epochs.  This is
true also for the Bouncing cDE model, which for low overdensity
thresholds appears practically indistinguishable from $\Lambda $CDM at
$z=0$, but shows a significant excess in the expected number of
massive halos at higher redshifts. On the other hand, at higher
overdensity thresholds the Bouncing cDE model can be easily
distinguished from $\Lambda $CDM even at $z=0$, as the degeneracy with $\sigma _{8}$
is broken and the halo mass function shows a significant enhancement
over the expected halo abundance at large masses.  We have
investigated the origin of this dependence of the deviation from
$\Lambda $CDM for the Bouncing cDE scenario, and found that the effect
is due to a significant increase of the average halo concentration at
low redshifts for such cosmology: starting from the same normalization
of the concentration-mass relation as $\Lambda $CDM at $z=2$, the
Bouncing cDE model produces concentrations that are twice as large as
in $\Lambda $CDM at $z=0$.

In a recent analysis \cite{Tarrant_etal11} computed the
linearly--extrapolated critical overdensity for spherical collapse,
$\delta_*$, for coupled DE models, with the purpose of including the
effect of coupling in the halo mass function. Clearly, including the
coupling effect in the computation of $\delta_*$ affects the predicted
halo abundance when using the expressions by \cite{Press74} (PS) or by
\cite{Sheth99} (ST). On the other hand, the fitting functions by
\cite{Jenkins01} (J01) and \cite{Tinker08} (T08) are not affected by
$\delta_*$ since their universality makes them to be only functions of
the r.m.s. fluctuation amplitude computed at the halo mass
scale. While it is beyond the scope of this paper to test the
predictions of PS and ST mass functions by accounting for DE coupling
on spherical collapse, our results clearly demonstrate that the
universal fitting functions by J01 and T08 can not be used to predict
mass and redshift dependence of halo counts with an accuracy of 10 per
cent or better. This is even more true when dealing with non--standard
cDE models, such as the EXP008e3 and SUGRA003 ones.
\ \\

To conclude, we have investigated the effects of interacting Dark
Energy models on the abundance of massive halos as a function of
redshift and mass, using different methods to identify halos and to
compute their mass.  Our results show that a clear degeneracy exists
between the coupling and the standard parameter $\sigma _{8}$, but
that both the redshift evolution and the detailed shape of the
high-mass tail of the halo mass function allow to break such
degeneracy at some level. Furthermore, we have shown that halo
concentrations are very mildly affected by the Dark Energy
interactions when the amplitude of density perturbations is normalized
at high redshifts (contrarily to what is found for a normalization at
$z=0$) with the only exception of the ``Bouncing'' coupled Dark Energy
scenario that shows a very rapid increase of halo concentrations with
respect to the fiducial $\Lambda $CDM model at recent epochs.  Our
study therefore provides a direct way to test interacting Dark Energy
models with present and future data on the abundance of massive
clusters as a function of redshift. A full exploitation of future
cluster surveys to constrain coupled DE models will however require
accurately calibrating corrections to expressions of the halo mass
function, whose universality has been tested with simulations only
within the $\Lambda$CDM framework.  Clearly, the observational
  measurement of the evolution of the cluster mass function alone
  could not provide in itself an incontrovertible test for the
  presence of cDE. This test should be complemented by other
  observational probes of large-scale structures. In this framework,
  the calibration of the HMF for cDE models presented here add an
  important piece of information that, in combination with other
  observational cobstraints, will enable to test and, possibly, to falsify the
  cDE scenario.

\section*{Acknowledgements}

The authors would like to thank Federico Marulli for valuable discussions.
Weiguang Cui acknowledges a fellowship from the European Commission's
Framework Programme 7, through the Marie Curie Initial Training
Network CosmoComp (PITN-GA-2009-238356).  MB is supported by the DFG
Cluster of Excellence ``Origin and Structure of the Universe'' and by
the TRR33 Transregio Collaborative Research Network on the ``Dark
Universe''. SB acknowledges partial financial support from the
PRIN-INAF-2009 Grant ``Towards an Italian Network for Computational
Cosmology'', by the PRIN-MIUR-2009 Grant ``Tracing the growth of
structures in the Universe'' and by the PD51-INFN Grant.

\bibliographystyle{mnras}
\bibliography{bibliography,paper}

\label{lastpage}

\end{document}